\documentclass[prl,amsmath,amssymb,twocolumn,showpacs,amsfonts,twocolumn]{revtex4}
\usepackage{graphicx}

\def \beq {\begin{equation}}
\def \eeq {\end{equation}}

\begin{document}
\title{Quantum Measurement Theory Explains the Deuteration Effect in Radical-Ion-Pair Reactions}
\author{A. T. Dellis and I. K. Kominis}
\email{ikominis@iesl.forth.gr}

\affiliation{Department of Physics, University of Crete, Heraklion
71103, Greece} \affiliation{Institute of Electronic Structure and
Laser, Foundation for Research and Technology, Heraklion 71110,
Greece}

\date{\today}

\begin{abstract}
It has been recently shown that radical-ion pairs and their reactions are a paradigm biological system manifesting non-trivial quantum effects, so
far invisible due to the phenomenological description of radical-ion-pair reactions used until now. We here use the quantum-mechanically consistent master equation
describing magnetic-sensitive radical-ion-pair reactions to explain experimental data [C. R. Timmel and K. B. Henbest, Phil. Trans. R. Soc. Lond. A
{\bf 362}, 2573 (2004); C. T. Rodgers, S. A. Norman, K. B. Henbest, C. R. Timmel and P. J. Hore, J. Am. Chem. Soc. {\bf 129} 6746 (2007)] on the
effect of deuteration on the reaction yields. Anomalous behavior of radical-ion-pair reactions after
deuteration, i.e. data inconsistent with the predictions of the phenomenological theory used so far, has been observed since the 70's
and has remained unexplained until now.
\end{abstract}
\maketitle The possible existence of non-trivial quantum effects
in biology \cite{davies} has fueled a tumultuous and ongoing
debate \cite{abbott}, as on the one hand effects associated with
quantum coherence are generally understood to be suppressed in the
typical biological/biochemical environment, on the other hand, it
is rightly assumed that nature must have found a way to utilize
the operational advantages offered by quantum physics, in
particular quantum coherence\cite{engel,fleming} and quantum
entanglement.

It has been recently shown \cite{kom} that a familiar biological
system, namely radical-ion pairs and their reactions, exhibits the
full spectrum of non-trivial quantum effects familiar from quantum
information science, namely quantum coherence, quantum jumps, the
quantum Zeno effect, and in principle quantum entanglement.
Radical-ion pairs play a fundamental role in a series of
biologically relevant chemical reactions, ranging from charge
transfer initiated reactions in photosynthetic reaction centers
\cite{boxer} to magnetic sensitive reactions abounding in the
field of spin-chemistry \cite{timmel}, and in particular in the
biochemical processes understood to underlie the biological
magnetic compass of several species having the ability to navigate
in earth's magnetic field \cite{schulten,ritz}.

\begin{figure}
\includegraphics[width=8 cm]{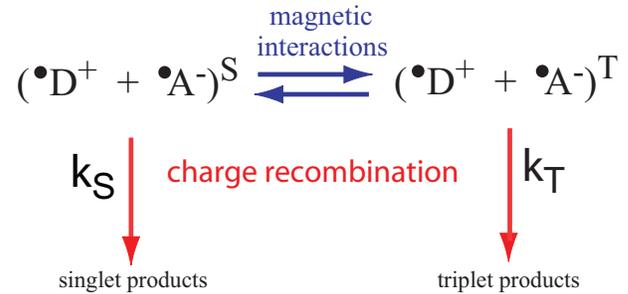}
\caption{Radical-ion-pair reaction dynamics: magnetic interactions
within the donor-acceptor molecule are responsible for the
singlet-triplet coherent mixing, while charge recombination into
singlet and triplet products, with respective rates $k_{S}$ and
$k_{T}$ removes molecular population from the singlet-triplet
subspace.} \label{f1}
\end{figure}

In Fig. \ref{f1} we depict a generic model for radical-ion-pair
reactions. Photoexcitation of a donor-acceptor molecule DA
followed by charge-transfer creates a radical-ion-pair with the
two unpaired electrons in the singlet state $(D^{+}$A$^{-})^{\rm
S}$. Magnetic interactions induced by the external magnetic field
and the internal hyperfine couplings of the unpaired electrons
with the molecule's nuclear spins bring about a coherent mixing of
$(D^{+}$A$^{-})^{\rm S}$ with $(D^{+}$A$^{-})^{\rm T}$, the
triplet radical-ion-pair. Both singlet and triplet radical-ion
pairs charge-recombine with rates $k_{S}$ and $k_{T}$ to singlet
and triplet products, respectively. This reaction forms a magnetic
sensor since the product yields depend on the external magnetic
field. However, radical-ion-pair reactions have so far been
described (see e.g. review \cite{steiner}) with a phenomenological
density matrix equation that (a) has by design not allowed the
true quantum nature of these reactions to be unveiled, and (b) has
led to several inconsistencies between theory and experiment. We
have recently put forward \cite{kom} the correct quantum dynamic description
of radical-ion-pair reactions following from quantum measurement
theory. In fact, there is a very similar system familiar in
condensed matter physics, namely coupled quantum dots, that is
treated with the same formalism, as described among others, in the
works of Milburn, Wiseman and co-workers \cite{mw}.
We will here use this quantum-mechanically consistent formalism to
account for quite a severe disagreement between experiments and
the phenomenological theory that has to do with deuteration. We
will first fully reproduce reaction data of the radical-ion-pair
Pyrene-Dymethylaniline (Py-DMA) and its deuterated versions, while
we also show that the phenomenological theory fails to account for
the same data. We will then use a simple theoretical model of a
radical-ion-pair to explain why the observed effects are
fundamentally different from what one would expect from the
phenomenological dynamics which in some cases is rather intuitive.
Using this simple model we will finally account for an older
observation of this "anomalous" effects of deuteration dating to
1979.

For completeness, we here reiterate the phenomenological as well as the new master equation describing the spin state evolution of the radical-ion-pair.
The spin Hilbert space of every radical-ion-pair consists of the two unpaired electron spins and any number of nuclear spins. This spin system is an
open quantum system due the reservoir states inducing charge recombination. The master equation describes both the unitary evolution of the spin state
due to magnetic interactions as well as the measurement evolution due to the reservoir states (vibrational excited states of the neutral recombined
molecule).
\begin{figure}
\includegraphics[width=8 cm]{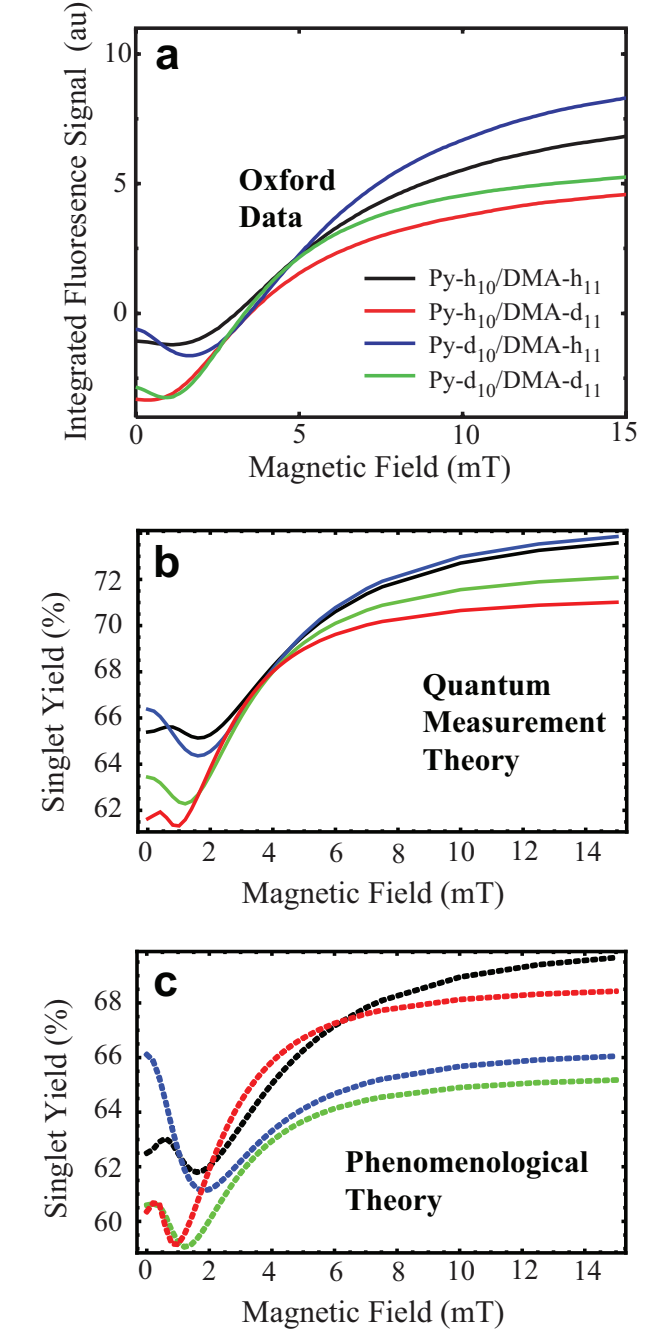}
\caption{(a) Experimental data from \cite{timmel,hore_data}. (b)
Reproduction of the data with quantum measurement theory equations
(\ref{eq:qev}) and (\ref{eq:jumps}). (c) Failure of
phenomenological theory (\ref{eq:cev}) to reproduce data.}
\label{f2}
\end{figure}
Until now the spin state evolution of a radical-ion-pair has been
described with the phenomenological master equation \beq
d\rho/dt=-i[{\cal H},\rho]-k_{S}(Q_{S}\rho+\rho
Q_{S})-k_{T}(Q_{T}\rho+\rho Q_{T})\label{eq:cev} \eeq where
$Q_{S}$ and $Q_{T}$ are the singlet and triplet state projection
operators, respectively. The first term in (\ref{eq:cev}) is the
unitary evolution due to the magnetic interactions, while the
second and third terms attempt to take into account population
loss out of the radical-ion-pair Hilbert subspace due to charge
recombination. It is these terms that are phenomenological and
that suppress by force the existing quantum-mechanical effects
like the quantum Zeno effect \cite{qZ}. On the other hand, the
actual quantum dynamic evolution of the radical-ion-pair spin density
matrix $\rho$ is given by the equations \beq d\rho/dt=-i[{\cal
H},\rho]-(k_{S}+k_{T})(Q_{S}\rho+\rho Q_{S}-2Q_{S}\rho
Q_{S})\label{eq:qev} \eeq where the second term describes the
measurement-induced evolution brought about by the singlet (rate
$k_S$) and triplet (rate $k_{T}$) recombination channel. Due to
the completeness relation $Q_{S}+Q_{T}=1$, both channels
effectively "measure" the observable $Q_{S}$ with a total
measurement rate $k_{S}+k_{T}$. The unconditional evolution
described by (\ref{eq:qev}) is interrupted by the
charge-recombining quantum jumps described by equations
(\ref{eq:jumps}) which give the probability of the singlet and
triplet recombination taking place within the time interval
between $t$ and $t+dt$, i.e.
\begin{align}
dP_{S}&=2k_{S}\langle Q_{S}\rangle dt\nonumber\\
dP_{T}&=2k_{T}\langle Q_{T}\rangle dt,\label{eq:jumps}
\end{align}\
The density matrix $\rho$ describing the spin state of the two
unpaired electrons and $n$ nuclear spins has dimension
$d=4\prod_{j=1}^{n}(2I_{j}+1)$, where $I_j$ is the nuclear spin of
nucleus $j$. Coming to the magnetic Hamiltonian ${\cal H}$, it is
composed of ${\cal H}_{\rm Z}$, the Zeeman interaction of the two
unpaired electrons (nuclear Zeeman interaction is negligible) with
the external magnetic field $\mathbf{B}=B\mathbf{\hat z}$, ${\cal
H}_{\rm hf}$, the hyperfine couplings of the electrons with the
surrounding nuclear spins \beq {\cal H}={\cal H}_{\rm Z}+{\cal
H}_{\rm hf}, \eeq with ${\cal H}_{\rm Z}=\omega(s_{1z}+s_{2z})$,
${\cal H}_{\rm
hf}=\sum_{i=1}^{2}\sum_{j=1}^{n}{\mathbf{s}_{i}\cdot\mathbf{A}_{ij}\cdot\mathbf{I}_{j}}$,
where the sum over $i$ is for the two electrons and the sum over
$j$ for the $n$ nuclear spins, with $\mathbf{A}_{ij}$ the
hyperfine coupling tensor of electron $i$ with nuclear spin $j$.

We will now reproduce recent experimental data by Hore and
co-workers \cite{hore_data,timmel} on the reaction of the
radical-ion-pair Py-DMA and its deuterated versions. In the
reproduction of the experimental data we will use a simple
radical-ion-pair model consisting of just two spin-1/2 nuclei. In
Figure \ref{f2}a we depict the fluorescence measurement of the
Py-DMA reaction and its deuterated versions. In Figure \ref{f2}b
we reproduced the data with the hyperfine couplings shown in Table
1, while for the same parameters, the prediction of the
phenomenological theory (\ref{eq:cev}) is shown in Figure \ref{f2}c.
The agreement of the former and the severe discrepancy of the
latter with the data is rather obvious.

\begin{figure}
\includegraphics[width=8 cm]{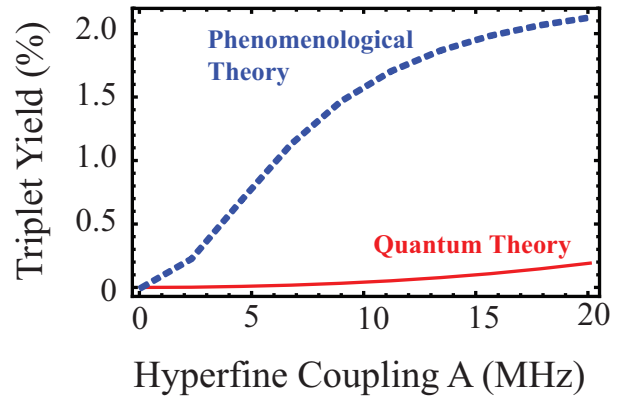}
\caption{Triplet yield of a one-nucleus (spin-1/2) radical-ion-pair as a function of the isotropic hyperfine coupling as calculated based on the quantum
theory (red continuous line) and the phenomenological theory (blue dashed line). The calculation was done for an external magnetic field of 0.5 g and the
singlet and triplet recombination rates were $k_{S}=20~\mu {\rm s}^{-1}$ and $k_{T}=0.5~\mu {\rm s}^{-1}$, respectively.}
\label{f3}
\end{figure}

Finally we will  resolve an old problem regarding deuteration that
clearly illustrates the qualitative understanding of radical-ion-pair reactions
that the phenomenological description fails to embody. To
illustrate the problem we use the simplest possible
radical-ion-pair model, that with just one spin-1/2 nucleus. We
calculate according to both quantum measurement theory
(\ref{eq:qev}) and (\ref{eq:jumps}) and phenomenological theory
(\ref{eq:cev}) the triplet yield of the reaction as a function of
the isotropic hyperfine coupling $A$, i.e. the magnetic Hamiltonian used for this calculations is just
${\cal H}=\omega(s_{1z}+s_{2z})+A\mathbf{s}_{1}\cdot\mathbf{I}$, with $I=1/2$. The result is shown in
Figure \ref{f3}. It is readily seen that the triplet yield stays
roughly constant according to the quantum measurement theory
\begin{table}
\caption{Hyperfine Couplings and Recombination Rates for the Data
Reproduction}
\begin{ruledtabular}
\begin{tabular}{|c|c|c|c|c|}
& ${\rm A_{Py}}$  &  ${\rm A_{DMA}}$    &   $k_{S}$  & $k_{T}$ \\
& (mT)                &        (mT)            &    ({\rm $\mu
s^{-1}$}) & ({\rm $\mu s^{-1}$})\\      \hline
Py-${\rm h}_{10}$ DMA-${\rm h}_{11}$ & 1.9   & 6.7   & 8.5      & 4.0  \\
Py-${\rm d}_{10}$ DMA-${\rm h}_{11}$ & 0.4   & 5.0   & 12.0     & 11.4  \\
Py-${\rm d}_{10}$ DMA-${\rm d}_{11}$ & 0.9   & 4.2   & 7.9      & 6.0  \\
Py-${\rm h}_{10}$ DMA-${\rm d}_{11}$ & 1.3   & 4.0   & 3.7      & 1.8  \\
\end{tabular}
\end{ruledtabular}
\end{table}
prediction, whereas it changes in a measurable way (a few \%) according to the phenomenological description. In fact,
the latter prediction is more intuitive, since it is based on the
following plausible description: In the case when the singlet
recombination rate is much larger than the triplet recombination
rate, as in this particular example we are considering, the
triplet yield should be proportional to the hyperfine coupling
$A$, as the probability to recombine through the triplet channel
should roughly scale as $A/k_{S}$, i.e. the singlet-triplet mixing
rate $A$ times the lifetime of the pair, $1/k_{S}$.

In Figure 4a we plot the evolution of the singlet projection
operator expectation value, $\langle Q_{S}\rangle$,  as calculated from the phenomenological density
matrix equation (\ref{eq:cev}), for two values of $A$, the
hyperfine coupling. In contrast, the predictions of quantum
measurement dynamics (\ref{eq:qev}) and (\ref{eq:jumps}) for these
two different values of $A$ are shown in Figure 4b and 4c,
together with the corresponding unitary evolution (no
recombination at all).

It is evident that for the same change in the hyperfine coupling, the relative change in
$\langle Q_{S}\rangle$ is in reality (quantum measurement theory) larger than predicted
by the phenomenological theory. However, due to the artificial structure of the latter (the phenomenological recombination terms) the absolute value
of $\langle Q_{S}\rangle$ changes substantially during the reaction. In contrast, the actual change of $\langle Q_{S}\rangle$ is in reality much smaller
exactly due to the projective nature of the recombination process that is embodied in the fundamental master equation (\ref{eq:qev}). Thus the triplet
reaction yield is in reality much less sensitive to the hyperfine coupling as naively and intuitively expected from the phenomenological understanding
of the reaction. In other words, the phenomenological master equation (\ref{eq:cev}) cannot account for the projective nature of the spin-conserving recombination channels, since it just describes a continuous disappearance of radical-ion-pairs due to the charge recombination process. On the contrary, as has been explained in \cite{kom}, the fundamental master equation (\ref{eq:qev}) takes into account the actual physical process that is going on: the recombination channels constitute a continuous measurement of the radical-pair's spin state, and as such, the latter is significantly affected.

The discrepancy that is visualized in Figure 3 is  what Blankenship and Parson observed in 1979 with the P870-I radical-ion-pair and its deuterated version. It is noted that deuteration is effectively equivalent to a change in the relevant hyperfine couplings \cite{blank}.
In fact, due to this severe inconsistency, these authors went as
far as to doubt the validity of the basic singlet-triplet
hyperfine mixing mechanism: "The identical triplet quantum yields
in H and D samples does not support the generally accepted idea
that hyperfine interactions are responsible for spin rephasing in
the P870 I radical pair". Mention of this problem has also been
made in the review \cite{boxer}.
\begin{figure}
\includegraphics[width=8 cm]{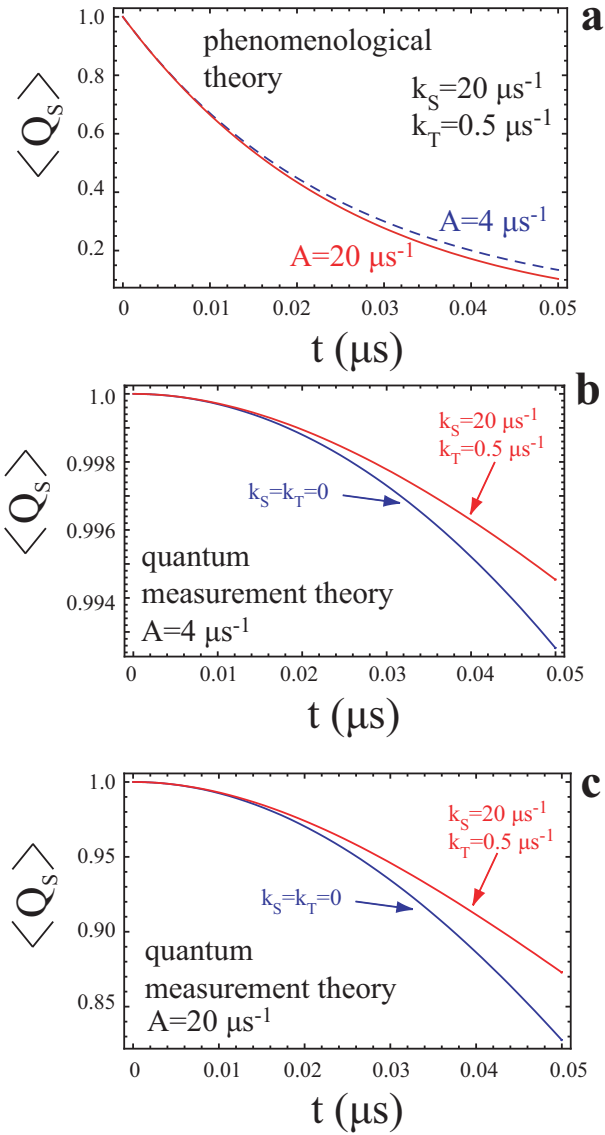}
\caption{(a) Evolution of $\langle Q_{S}\rangle$ based on the
phenomenological theory for a time scale on the order of the
duration of the reaction, $1/k_{S}=0.05~\mu {\rm s}$ and for two
different values of the hyperfine coupling $A$ (b) Evolution of
$\langle Q_{S}\rangle$ based on quantum measurement theory with
and without the recombination terms, for $A=4~\mu {\rm s}^{-1}$.
(c) Evolution of $\langle Q_{S}\rangle$ based on quantum
measurement theory with and without the recombination terms, for
$A=20~\mu {\rm s}^{-1}$.} \label{f4}
\end{figure}

In summary, we have shown that the quantum measurement dynamics
present in radical-ion-pair reactions firstly explain recent
experimental data and secondly resolve a long-standing discrepancy
between the phenomenological description of these reactions that
has been used until now and experimental observations. Radical-ion
pairs and their reactions represent the first biological system
where fundamental concepts of quantum mechanics and quantum information theory are
fruitfully applied. The biological significance, if any, of this
fact is a rather exciting question that surfaces from the above
described change in out fundamental understanding of
radical-ion-pair reactions.

\end{document}